\documentclass[12pt,prd,aps,showpacs,superscriptaddress,nofootinbib,amsmath,amssymb]{revtex4-1}
\usepackage{graphicx}% Include figure files
\usepackage{dcolumn}% Align table columns on decimal point
\usepackage{bm}% bold math
\usepackage{lipsum}
\usepackage{multirow}
\usepackage{adjustbox}
\usepackage{textcomp}
\graphicspath{{./eps/}}

\begin{document}
\title{Recent status of the understanding of neutrino-nucleus cross section}

\author{H. \surname{Haider}}
 \email{huma.haider8@gmail.com}
\affiliation{Department of Physics, Aligarh Muslim University, Aligarh-202002, India}

\author{M. Sajjad \surname{Athar}}
\email{sajathar@gmail.com}
\affiliation{Department of Physics, Aligarh Muslim University, Aligarh-202002, India}

\author{S. K. \surname{Singh}}
% \email{rafi.alam.amu@gmail.com}
\affiliation{Department of Physics, Aligarh Muslim University, Aligarh-202002, India}

\begin{abstract}
In this work we have presented current understanding of neutrino-nucleon/nucleus cross sections in the few GeV energy region relevant for a precise
determination of neutrino
oscillation parameters and CP violation in the leptonic sector. In this energy region various processes like  quasielastic and inelastic production of single and 
multipion production, coherent pion production, kaon, eta, hyperon production,
associated particle production as well as deep inelastic scattering processes contribute to the neutrino event rates.
%\keywords{neutrino, quasielastic, inelastic, DIS}
\end{abstract}
%\Large
\maketitle              % typeset the title of the contribution
\section{Introduction}

The neutrino-nucleus($\nu$-A) cross sections are important input to the systematics of analysing $\nu$ oscillation experiments.
The measured events are a convolution of energy-dependent neutrino flux and $\nu$-A cross section. 
For precise measurements, there are two challenging tasks, (i) good knowledge of neutrino
fluxes and (ii) well understood nuclear medium effects(NME) in the entire region of neutrino energy 
spectrum. In this presentation, we have considered the following neutrino interactions:
\begin{eqnarray*}
\rm{QE:}~~\nu_l + n &\rightarrow & l^- + p \nonumber\\
\rm{\pi:}~~\nu_l + N &\rightarrow & l^- + \pi^i + N^\prime;~i=+, ~0~ or~ -;~N,N^\prime=p~ or~ n  \nonumber\\
\rm{K:}~~\nu_l + N &\rightarrow & l^- + K^i + N^\prime;~i=~+~ or~ 0  \nonumber\\
\rm{\eta:}~~\nu_l + N &\rightarrow & l^-  + \eta + N^\prime  \nonumber\\
\rm{AP:}~~\nu_l + N &\rightarrow & l^- + Y +   K^i; ~Y=~\Lambda, \Sigma   \nonumber\\
\rm{DIS:}~~\nu_l + N &\rightarrow & l^-  + X;~ X=~jet~ of~ hadrons.
\end{eqnarray*}
on nucleon and nuclear targets. Similar reactions take place with $\bar\nu_l$. 
It is estimated that due to NME the cross sections
have an overall uncertainty of 20-25$\%$\cite{Katori:2016yel,Morfin:2012kn}. We present here a short review of current understanding of $\nu-A$ cross section. 

\section{Quasielastic(QE) reactions}
 
\begin{figure}[t]
    \includegraphics[height=4.8cm,width=4.2cm]{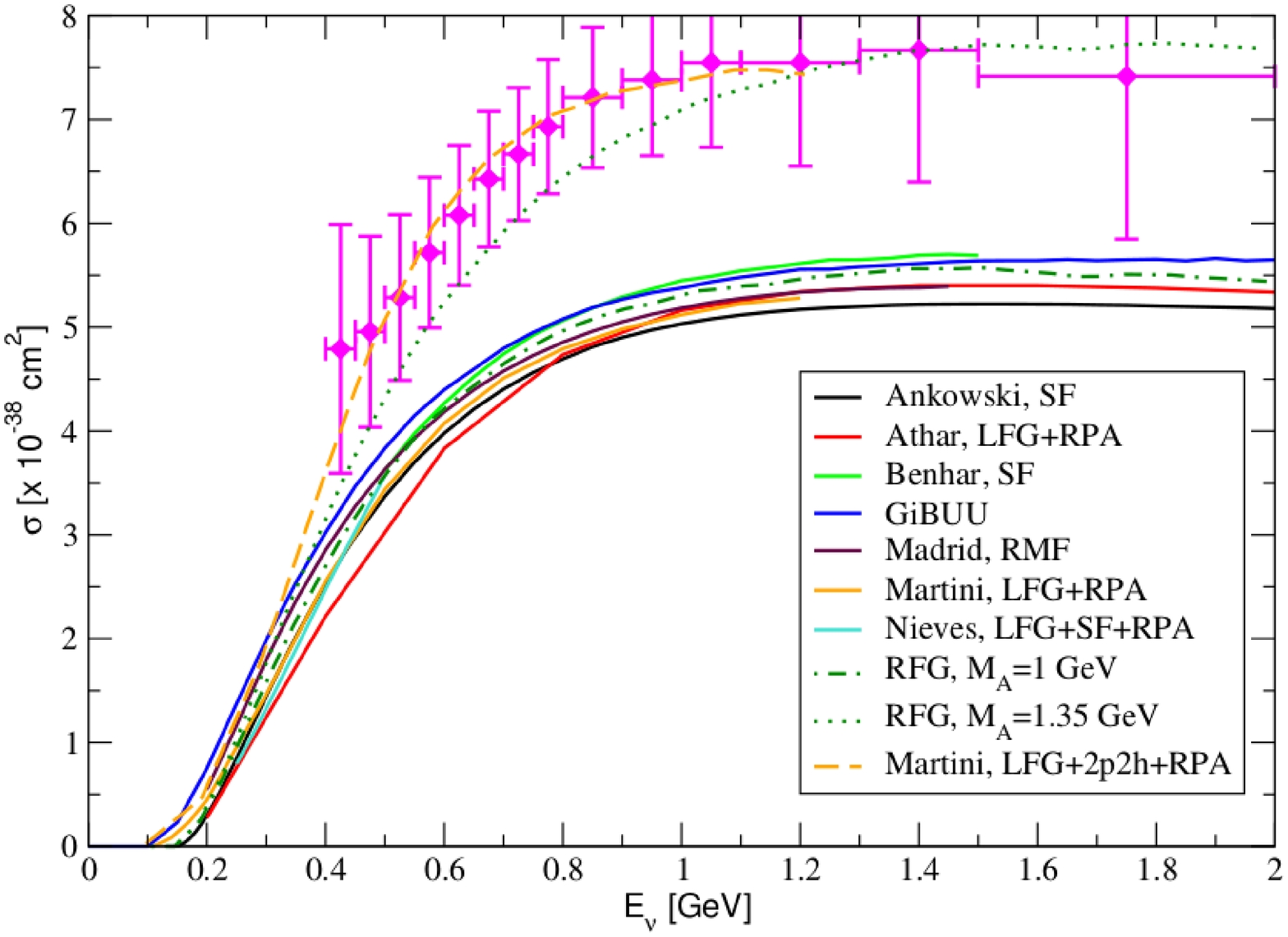}
        \includegraphics[height=5cm,width=4.2cm]{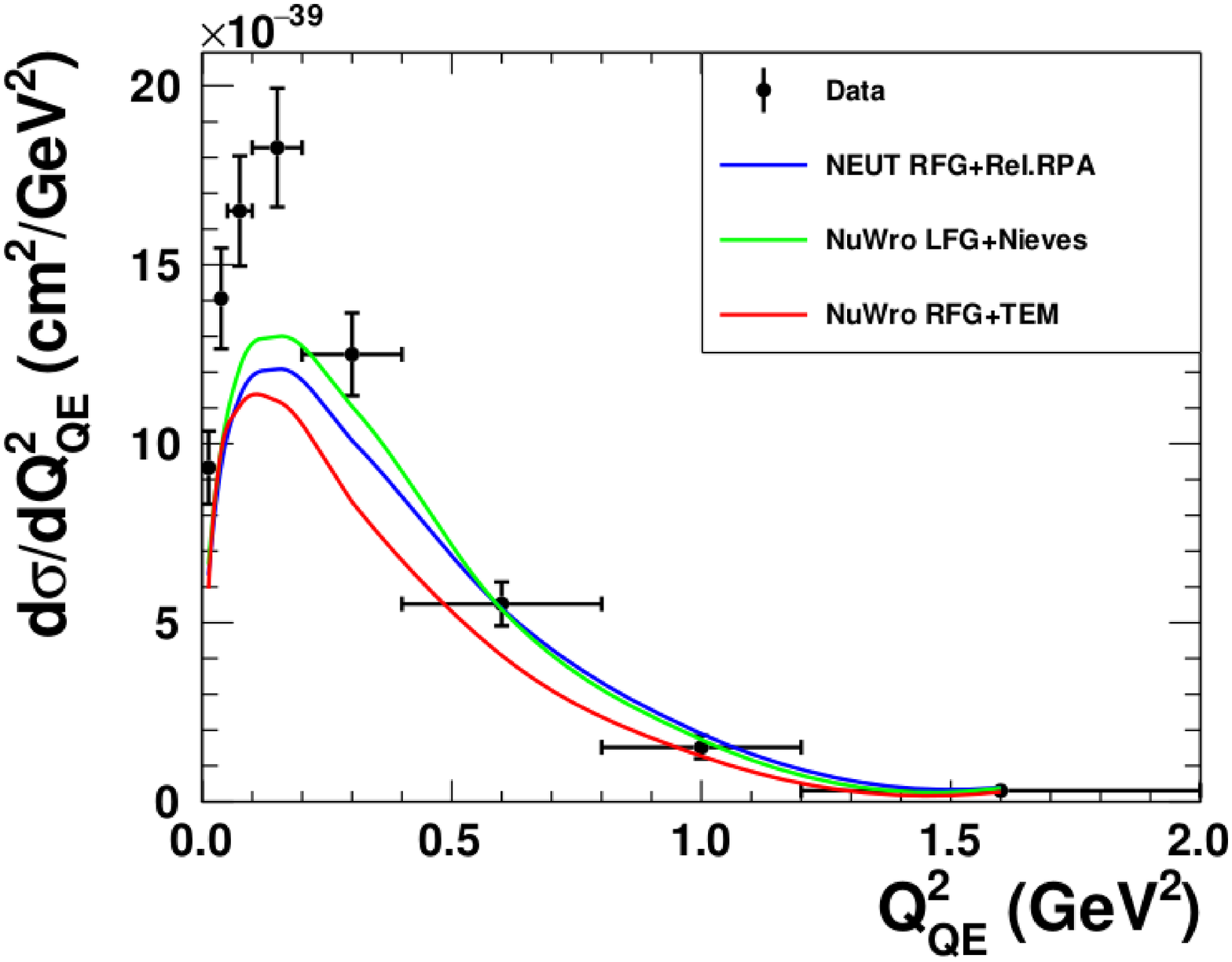}
    \includegraphics[height=5cm,width=4.2cm]{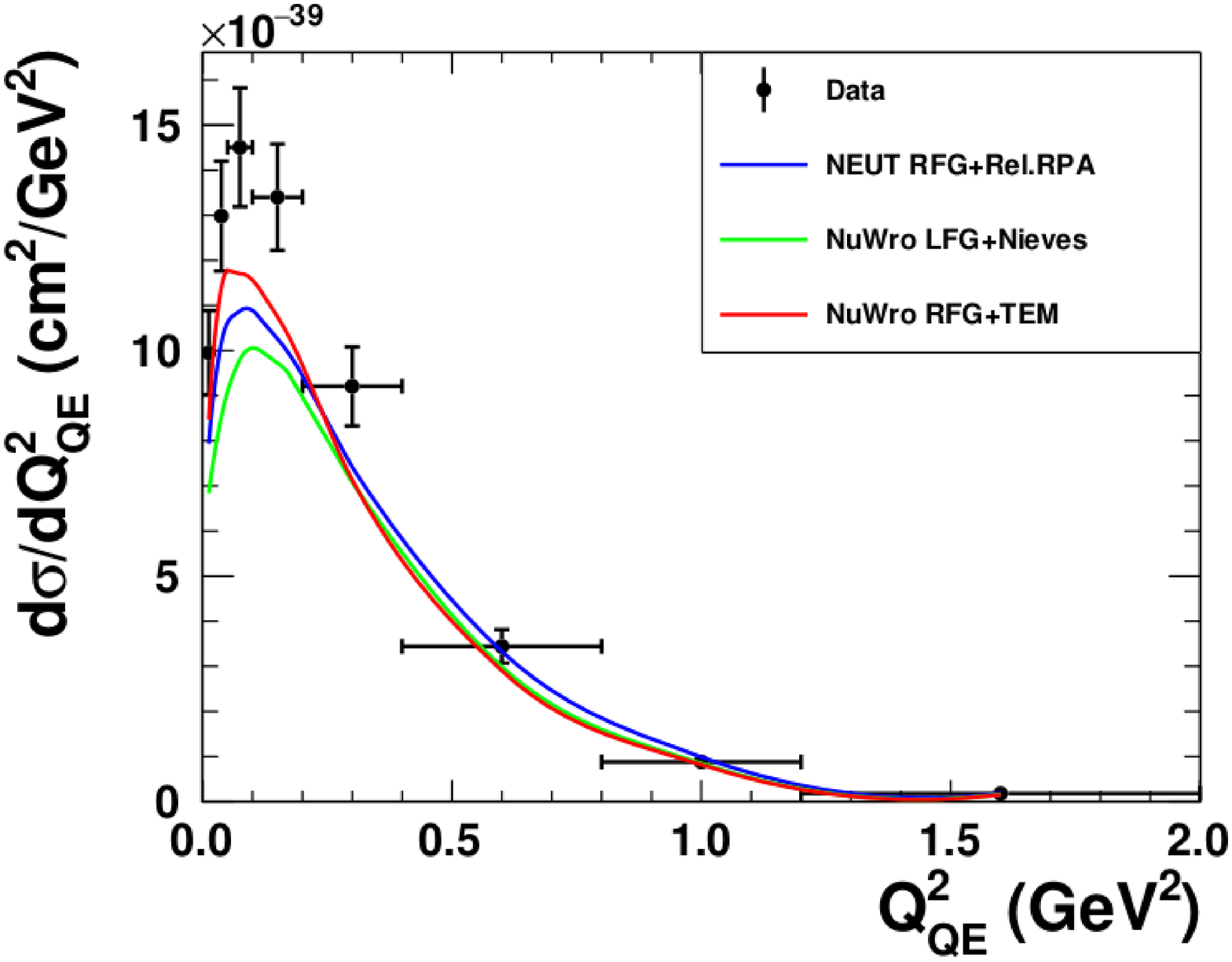}
    \caption{(Left) $\sigma_{CCQE}$ vs $E_{\nu_\mu}$ in $^{12}C$ calculated within several models along with the MiniBooNE data~\cite{Katori:2016yel}. 
    (Mid and Right) Comparison of the best fit MC predictions in NEUT and NuWro to MINERvA CCQE data. The clear difference in the normalisation between 
    the MC and the experimental data arises from the MINERvA data placing a much stronger constraint on the cross-section shape than its normalisation~\cite{Stowell:2016exm}.}
    \label{fig1}
\end{figure}

Most of the present Monte Carlo generators use
 relativistic Fermi gas model(FGM) given by  
Smith and Moniz\cite{Smith:1972xh} to analyse the experimental results on $\nu-A$ cross section. However, other variants of FGM like the model used by 
 Aligarh~\cite{Singh:1993rg,Athar:1999fu} and Valencia~\cite{Nieves:2004wx} groups, where local FGM 
 with long range correlations of the particle-hole excitations in the nuclear medium were included
 or the calculations where nucleon spectral functions are used to include the initial state interactions\cite{Benhar:2014qaa}.
  First high statistics experimental results from MiniBooNE~\cite{AguilarArevalo:2007ab}, and later K2K\cite{Gran:2006jn} and
 MINOS\cite{Dorman:2009zz} could not be explained by using Smith and Moniz\cite{Smith:1972xh} model of QE reactions using the standard values of 
 weak vectors and axial vector form factors. These experiments required a value of $M_A$ which was much larger than the world 
 average value of 1.026GeV. It was shown by Martini et al.\cite{Martini:2013sha} and later Nieves et al.\cite{Nieves:2013fr},
 the importance of two particle-two hole(2p-2h)
 contribution(which is basically multi-nucleon correlation effect). The inclusion of 2p-2h to the CCQE results obtained using FGM with RPA effect 
 gives satisfactory explanation of the data with $M_A=1.03$ GeV which are consistent with the values reported from
 NOMAD\cite{Lyubushkin:2008pe} and MINERvA\cite{Wilkinson:2016wmz} that corresponds to high $\nu$ energies. It is likely that these 2p-2h contributions
 are important in the region of low energies and are small at higher energies.
 The results are shown in Fig.1(left). Recently Stowell et al.\cite{Stowell:2016exm} compared the MINERvA CCQE data with two different MC generators NEUT
 and NuWro, the results of which have been shown 
 in Fig.\ref{fig1}(mid and right).
 It may be seen that the data and the results from the MC generators are not in agreement particularly in the low $Q^2$ region.   
 Therefore, better theoretical models to understand NME in $\nu$-A scattering process for analysing CCQE events are needed.
\begin{figure}[t]
    \includegraphics[height=4.7cm,width=4.2cm]{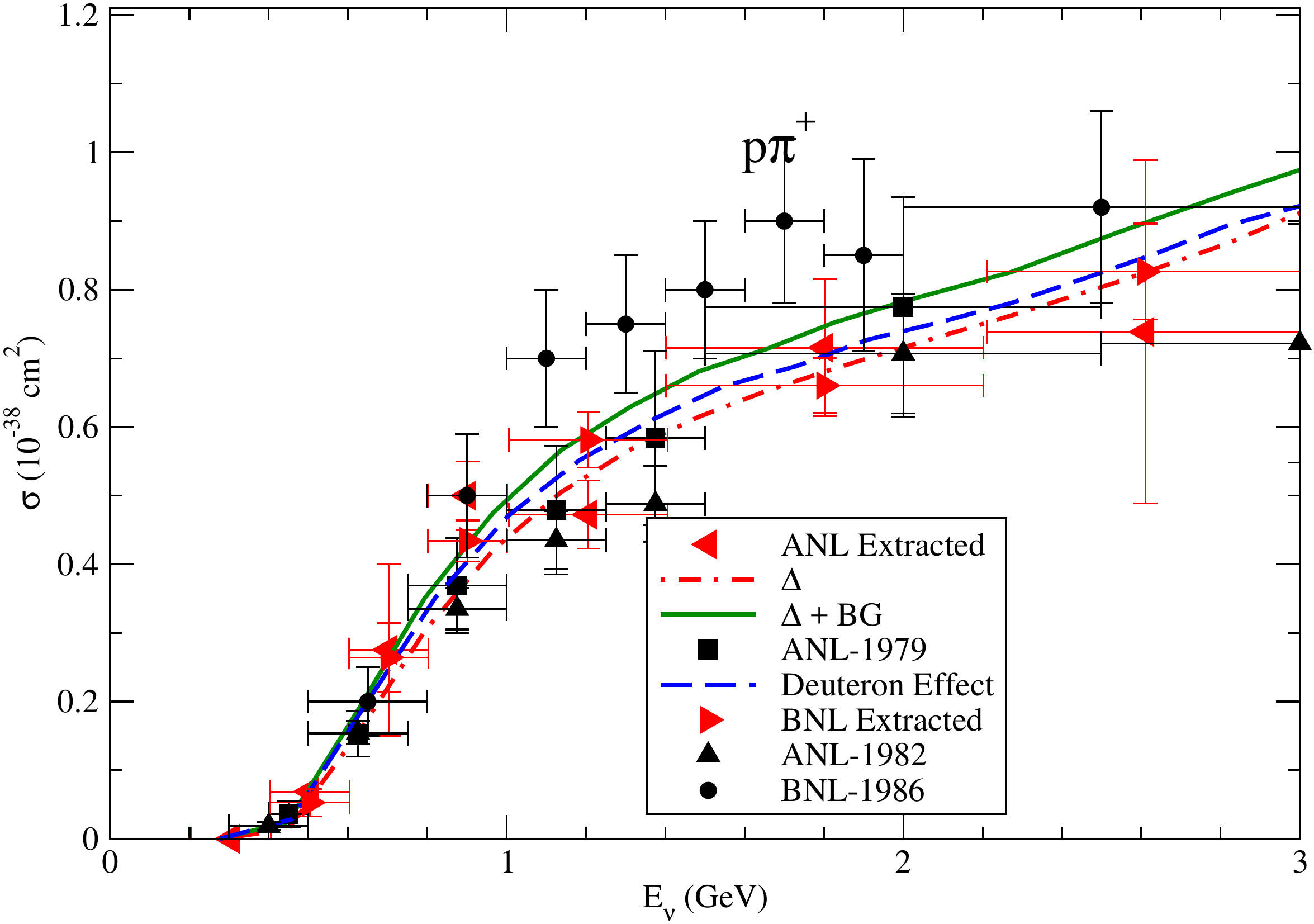}
        \includegraphics[height=5cm,width=4.3cm]{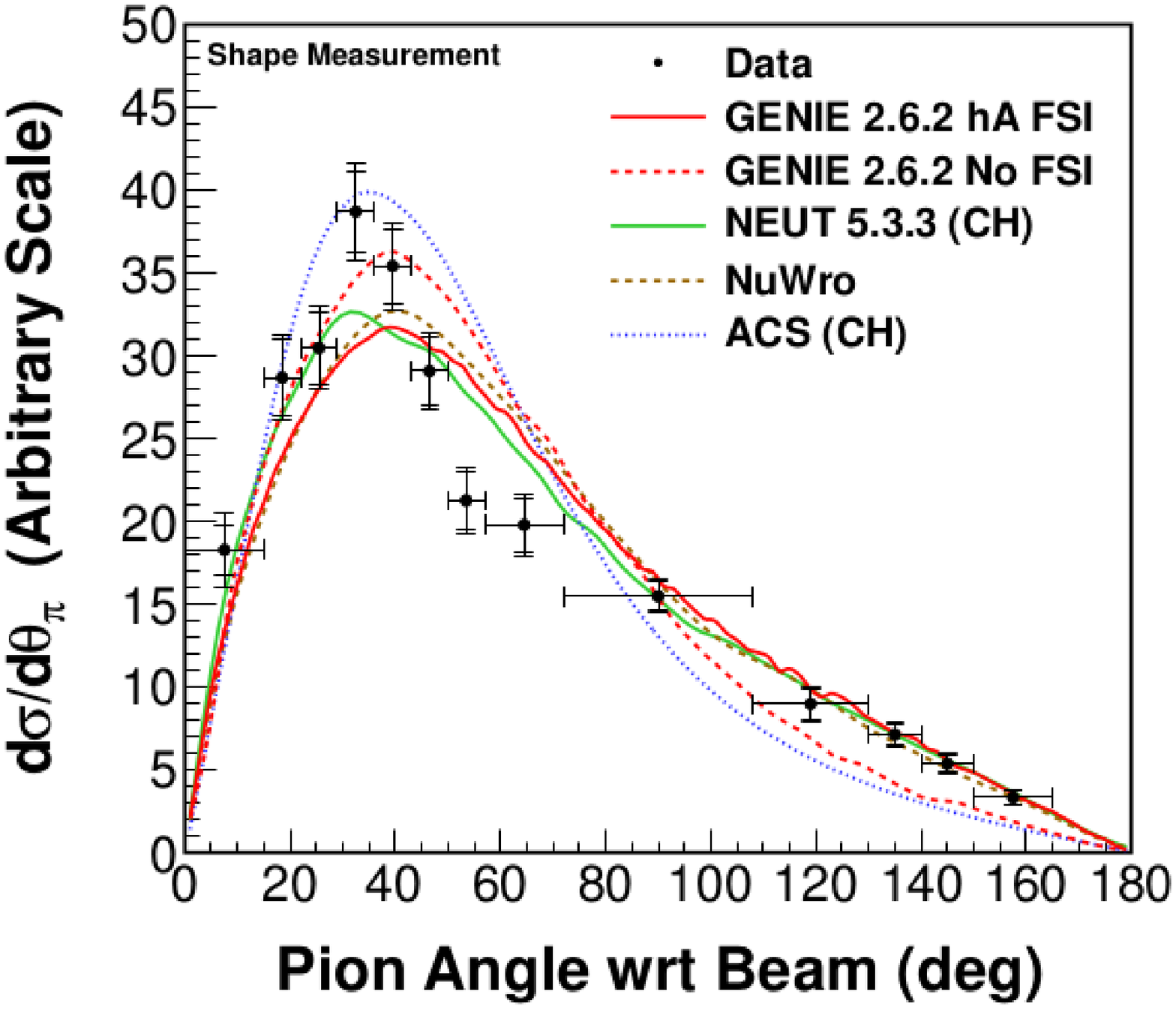}
            \includegraphics[height=5cm,width=4.2cm]{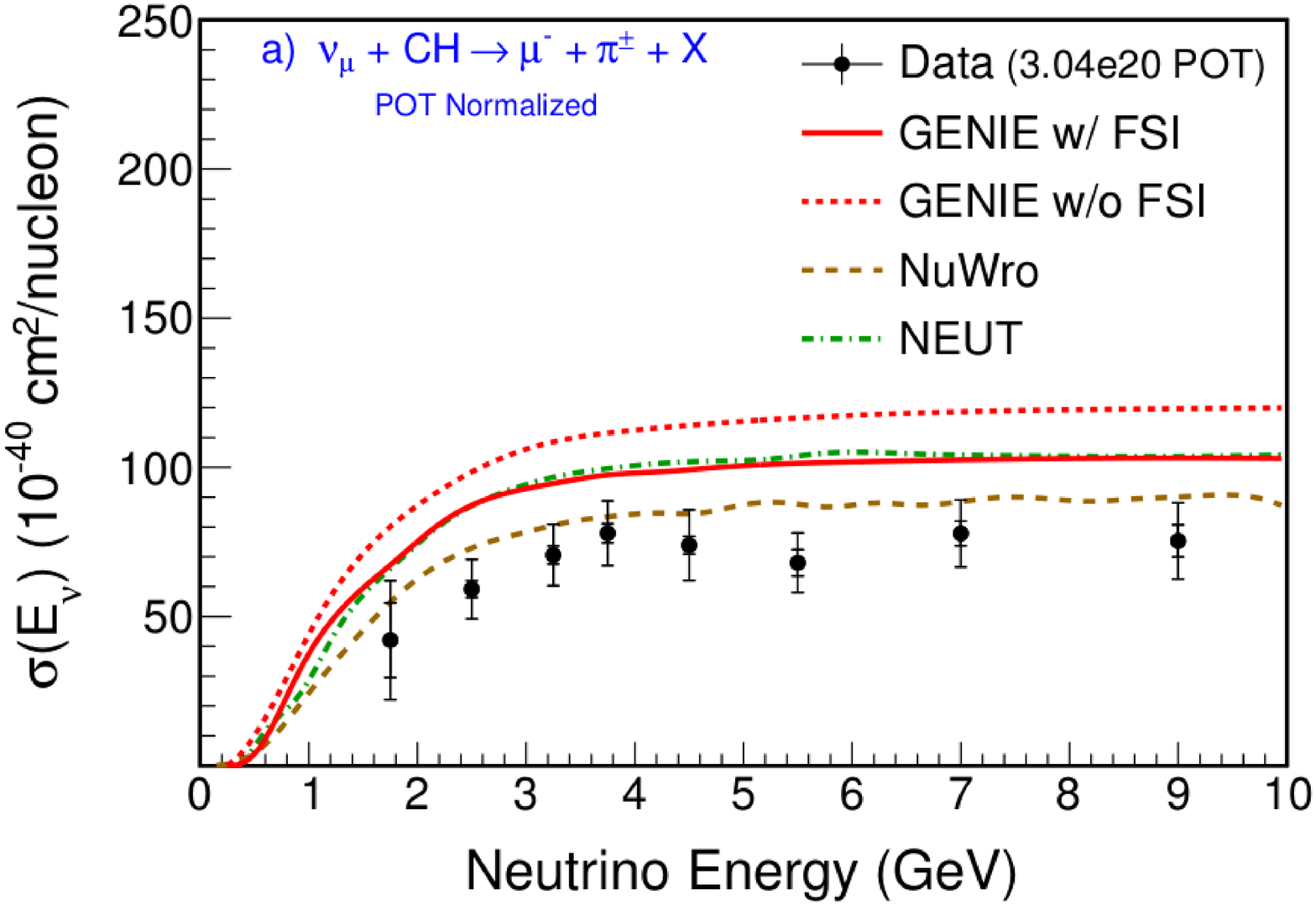}
            \caption{(Left) $\sigma_{CC1\pi}$ vs $E_{\nu_\mu}$ for $\nu_{\mu}   p \rightarrow \mu^{-}   p   \pi^{+}$ process. 
 Experimental results are reanalyzed data points of ANL and BNL experiments. For details please see Ref.~\cite{Alam:2015gaa}.
(Mid) Pion angular distribution in several models along with the MINERvA data~\cite{Eberly:2014mra}.
(Right) MINER$\nu$A data of CC1$\pi$ production and
 comparison with the different MC generators~\cite{McGivern:2016bwh}.}
\label{fig2}
\end{figure}

\section{Single $\pi$ production}

The experimental data on weak pion
production in $\nu$-A scattering in the experiments performed at 
 MiniBooNE, 
SciBooNE 
 and more recently from MINER$\nu$A 
 collaboration 
 have highlighted the inadequacy of our present 
 understanding of nuclear medium and final state interaction effects. In addition there is lack of consensus in the theoretical modeling of basic reaction mechanism of $\nu(\bar\nu)$ induced pion production 
 from free nucleon, specially concerning the role of background terms and higher resonances. By including the contribution of background terms and higher 
 resonances along 
 with the dominant $\Delta$ resonance, Aligarh group~\cite{Alam:2015gaa} has studied weak charged and neutral current induced single pion production from 
 nucleons and 
 fitted the reanalysed ANL and BNL data~\cite{Wilkinson:2014yfa}, and the results are shown in Fig.\ref{fig2}(left). It was concluded that the best description of 
 the reanalyzed experimental 
 data of ANL and BNL experiments is obtained 
 when we take $C_5^A(0)$=1.0 and $M_A$=1.026GeV for N$-\Delta$ axial vector transition current form factor $C_5^A(Q^2)$.
 In an another work by Valencia group~\cite{Alvarez-Ruso:2015eva}, implementing unitarity using Watson's theorem and applying it to fit
 old and reanalysed ANL and BNL data, the best fit was found with $C_5^A(0)=1.12\pm0.11$ and $C_5^A(0)=1.14\pm0.07$ respectively, and $M_A$=$0.954\pm 0.063$GeV.
  
  In the case of single $\pi$ production from nuclear targets, the dominant contribution from the $\Delta$ resonance is suppressed due to modifications of 
  mass and width of the $\Delta$ propagator in the nuclear medium~\cite{Athar:2007wd}.
   Moreover, the produced pions undergo final state interaction(FSI) with the residual nucleus which further reduces the pion production.
 In Fig.\ref{fig2}(mid), the results for CC1$\pi$ production 
 in $^{12}C$ obtained by different theoretical groups are shown along with the MINER$\nu$A data~\cite{Eberly:2014mra}.  
 Recently McGivern et al.\cite{McGivern:2016bwh} have published MINER$\nu$A data of CC1$\pi$ production and
 compared the results with different theoretical approach incorporated
 in MC generators(Fig.\ref{fig2} right). It may be clearly observed that there is 
 a discrepancy between the theoretical and experimental results which need further study.
\begin{figure}[t]
    \includegraphics[height=12cm,width=12cm]{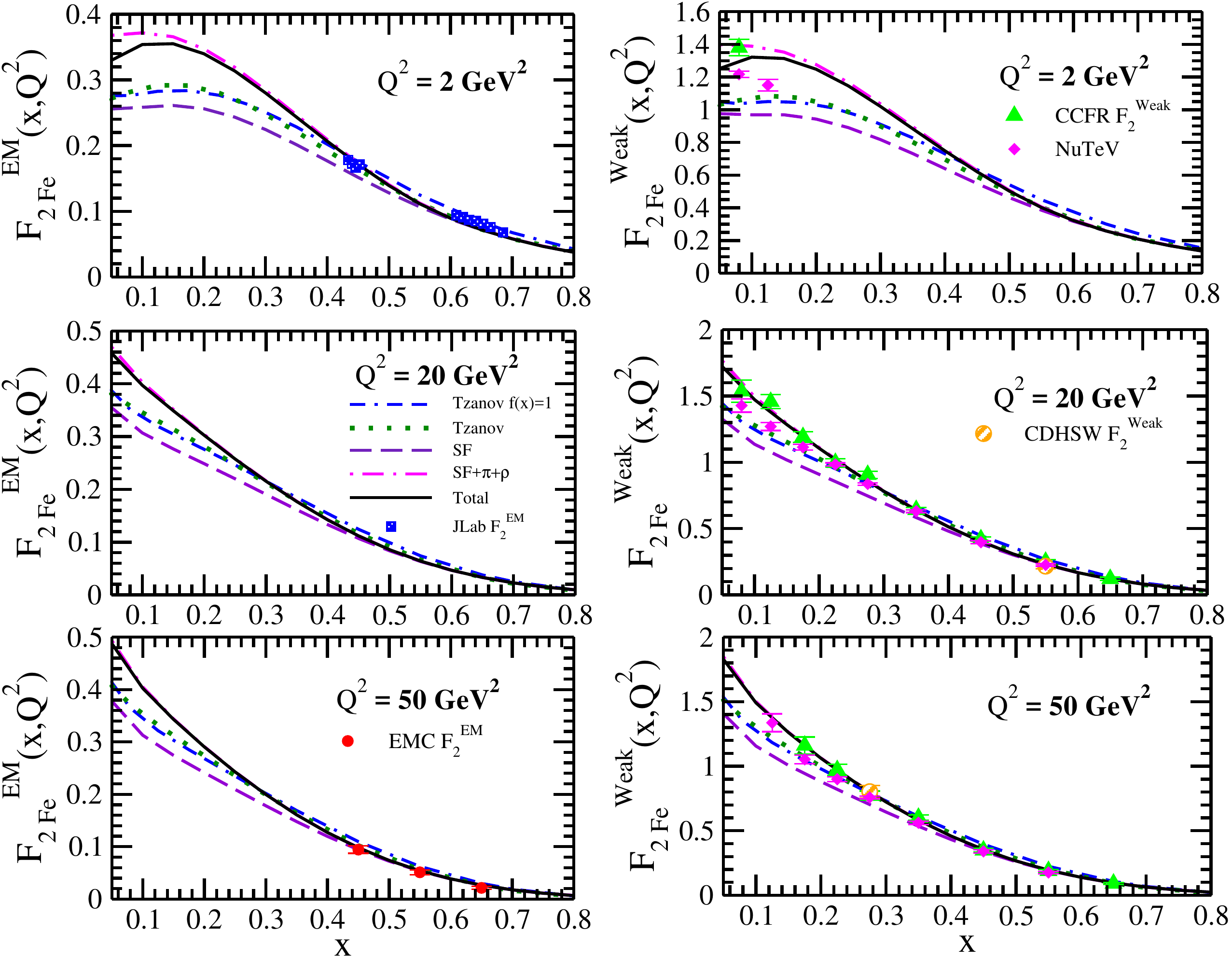}
    \caption{Results of EM(Left panel) and Weak(Right panel) nuclear structure functions in $^{56}Fe$(isoscalar) obtained 
 using spectral function(long dashed line), including mesonic 
contribution(dashed-dotted line), full model(solid line). The details may be seen at Ref.~\cite{Haider:2016zrk}.}
 \label{dis}
\end{figure}
There is another mode of pion production from nuclear target known as coherent pion production where a nucleus after the interaction remains in the ground state and all the energy is transferred to the 
outgoing pion. The work by the Aligarh group\cite{Singh:2006br} and the Valencia group~\cite{AlvarezRuso:2007tt} have shown that the contribution from coherent pion production is 
highly suppressed due to NME and is 2-3$\%$ of the total pion production. Besides 1$\pi$ production
there can be 2$\pi$ or multipion production for which details can be found in Ref.~\cite{Hernandez:2007ej}. 

\section{Deep-inelastic scattering (DIS)}

 There is no sharp kinetic region to distinguish the onset of the DIS region 
from the resonance region but the region $ W \ge 2.0 $ GeV and $Q^2 \ge 1.0$ GeV$^2$ is considered to be 
DIS region. 
At high energy and $Q^2$, the inclusive DIS cross sections are usually expressed in
terms of the structure functions
which are derived in terms of quark PDFs using the methods of perturbative QCD. These structure functions are experimentally determined from DIS experiments
on nucleon and nuclear targets. The observation of EMC effect has led to the presence of strong NME in DIS region.
Phenomenologically NME on quark PDFs and nucleon SF have been determined by many groups~\cite{Kovarik:2012zz}. There is considerable theoretical work on 
 studying NME in structure
functions~\cite{Malace:2014uea}. Recently a comparative study of NME in  $F_{2A}^{EM}(x,Q^2)$ and $F_{2A}^{Weak}(x,Q^2)$ has been done by 
Aligarh group~\cite{Haider:2011qs,Haider:2016zrk} and 
Kulagin and Petti~\cite{Kulagin:2007ju}. 

The Aligarh group has studied NME in  structure functions for moderate as well as heavy nuclear targets. We construct a relativistic nucleon spectral function 
for a nucleon in an interacting Fermi
sea within a field theoretical formalism which uses nucleon
propagators in the nucleus. 
The spectral function takes into account binding energy, Fermi motion and nucleon correlations. 
 Other effects like target mass correction, next to leading order correction,
 pion and rho mesons cloud contributions, shadowing and antishadowing corrections are also included in numerical calculations of structure functions. 
 In Fig.\ref{dis}, we show the curves for $F_{2A}^{EM,Weak}(x,Q^2)$ obtained using the spectral
 function only, also including the mesonic contribution, 
 and finally using the full model which also includes shadowing. 
 The use of spectral function leads to a reduction in $F_{2A}^{EM}(x,Q^2)$ as well as in $F_{2A}^{Weak}(x,Q^2)$
 nuclear structure functions as compared to the free nucleon case. 
 The inclusion of mesonic contributions from pion and rho mesons leads to an enhancement in these structure functions at low and mid values of $x$.
  The inclusion of shadowing effects further 
   reduces these structure functions and are effective in low region of $x~(~<~0.1)$. 
     \begin{figure}[t]
        \includegraphics[height=5cm,width=4cm]{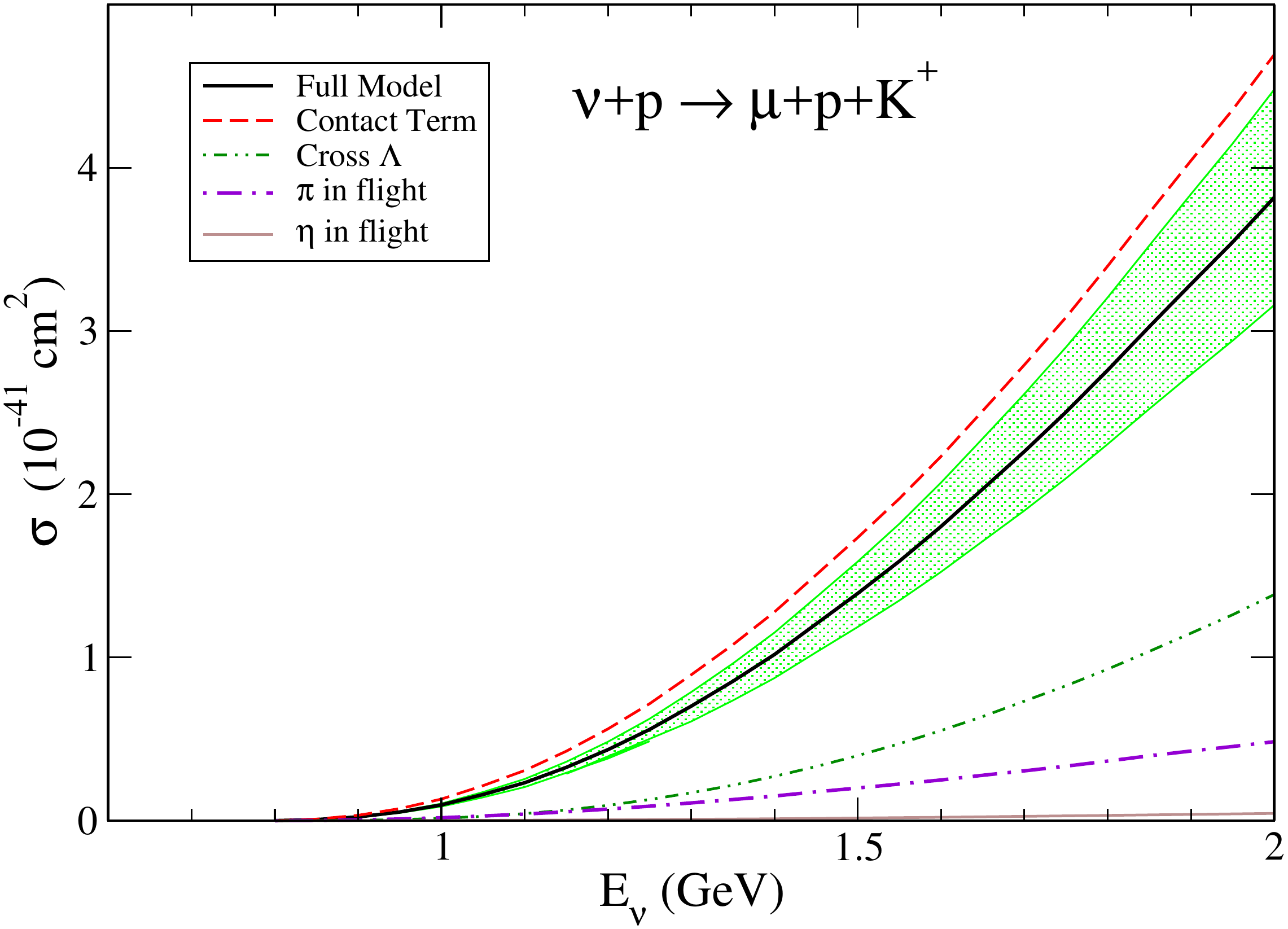}
    \includegraphics[height=5cm,width=4cm]{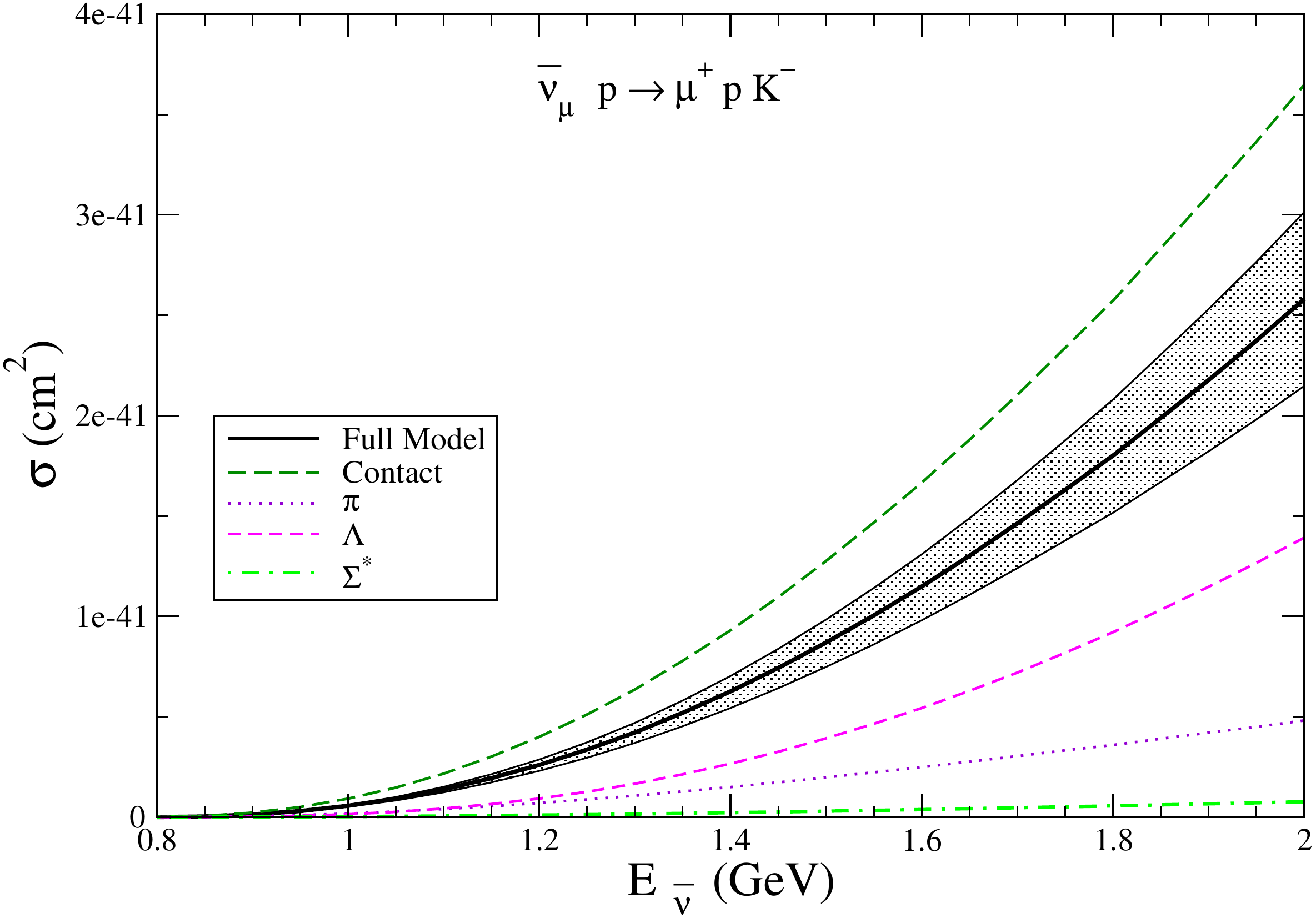}
              \includegraphics[height=5cm,width=4cm]{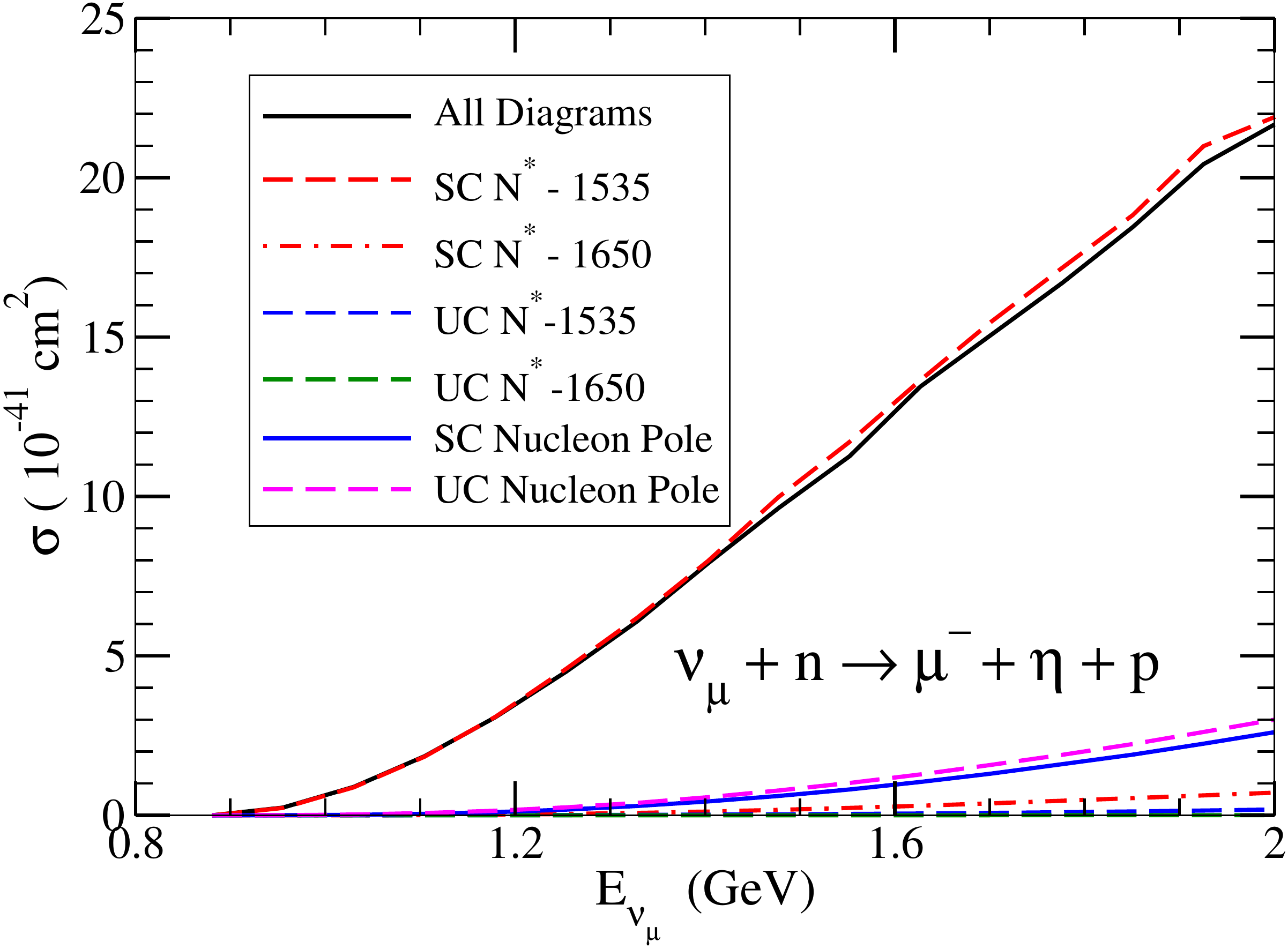}
               \caption{Cross section for $\nu$ induced (left) $K^+$, (mid) $K^-$ and (right) $\eta$ production processes.
              The amplitude for the hadronic current gets contribution from various Feynman diagrams viz. direct, cross terms, 
              contact diagram, pion pole,and 
 pion/eta in flight. We have also included $\Sigma^*$(1385) resonance for antikaon production. For the $\eta$ production the major contribution 
 is from s-channel $S_{11}(1535)$ resonance followed by u-channel contribution and from the Born diagrams.}
 \label{strange}   
\end{figure}

 \section{$|\Delta S| =0 $ and $|\Delta S| = 1 $ processes off nucleon}
 
  The other inelastic processes like $K$, $\eta$ and AP production have also been discussed briefly. 
 The strange particles are produced by both $|\Delta S| =0 $ and $|\Delta S| = 1 $ processes. 
At the neutrino energies of $\sim$1GeV it is the single hyperon(by $\bar \nu$) or single kaon($K/\bar K$) that are produced by 
$|\Delta S| = 1 $ reaction mechanism while $\eta$ meson and associated production of kaon are induced
by the $|\Delta S| = 0 $ weak currents. 
The reaction cross sections are smaller than the pion production due to Cabibbo suppression in $|\Delta S| = 1 $ process. It has been shown by Aligarh group
~\cite{RafiAlam:2010kf,Alam:2015zla,Singh:2015ama,Akbar:2016awk}  
that for the precise determination of neutrino oscillation parameters and 
to estimate the background for a study of nucleon decay searches, these reactions are important.
In Fig.\ref{strange}, we have presented the results for CC induced $K$, ${\bar K}$ and
$\eta$ production cross sections. The details are given in Ref.~\cite{RafiAlam:2010kf}-\cite{Akbar:2016awk}.

 \end{document}